\documentclass[aps,twocolumn,pre,showpacs,color,psfig,epsf]{revtex4}
\usepackage{amsmath}
\usepackage{color}
\usepackage{amsfonts}
\usepackage{epsf}
\usepackage{graphicx}
\begin{document}

\title{Facilitated diffusion on confined DNA}


\author{G. Foffano$^1$, D. Marenduzzo$^1$, E. Orlandini$^2$}
\affiliation{$^1$SUPA, School of Physics and Astronomy, University of Edinburgh, Mayfield Road, Edinburgh, EH9 3JZ, UK;\\
$^2$Dipartimento di Fisica e  Sezione INFN, Universit\`a di Padova, Via Marzolo, 31021 Padova, Italy}

\begin{abstract}
In living cells, proteins combine 3D bulk diffusion and 1D sliding along the DNA to reach a target faster. This process is known as facilitated diffusion, and we investigate its dynamics in the physiologically relevant case of confined DNA. The confining geometry and DNA elasticity are key parameters: we find that facilitated diffusion is most efficient inside an isotropic volume, and on a flexible polymer. By considering the typical copy numbers of proteins {\it in vivo}, we show that the speedup due to sliding becomes insensitive to fine tuning of parameters, rendering facilitated diffusion a robust mechanism to speed up intracellular diffusion-limited reactions. The parameter range we focus on is relevant for {\it in vitro} systems and for facilitated diffusion on yeast chromatin.
\end{abstract}

\maketitle

\section{Introduction}
The binding of proteins to DNA or chromatin is at the basis of gene regulation, as for instance transcription factors stick to a gene along the DNA to promote or inhibit the subsequent binding of RNA polymerases, and hence to down or upregulate its transcription~\cite{Alberts,Voituriez,Voituriez2}.

In 1970 Riggs found that the association rate of the {\it lac}-repressor 
in {\it E.coli} is ~$10^{10}M^{-1}s^{-1}$, two orders of magnitude larger 
than it would be if the reaction time were dominated by the 3-dimensional (bulk) 
diffusion of the proteins through cytoplasm~\cite{Riggs}. 
How can proteins move around so quickly in the cell? 
An intriguing mechanism, alternative to bulk diffusion, was proposed by Berg and von Hippel 
in~\cite{Berg_I}. They conjectured that proteins may bind non-specifically to DNA and diffuse along it -- the linear portion $l_s$ of DNA which is explored before the protein detaches from it is known as the {\it sliding length}. 
DNA looping may also allow hops and jumps of the protein between different sites on the DNA~\cite{metzler,def_jump_hop}. The expression {\it facilitated diffusion} indicates all the different mechanisms through which proteins exploit the presence of DNA in the cell in order to reduce 
the {\it mean search time} (hereon MST), namely the average time that one protein needs to reach its target. 
By considering a protein performing a very large number of slides, attachments and detachments, it is possible to theoretically estimate the value of the MST as a function of the sliding length $l_s$  (see e.g.~\cite{Marko,Voituriez,Voituriez2}). 
The theory leads to the intriguing result that there should be an optimal 
sliding length for which the mean search time is minimum, and 
significantly smaller than in the bulk diffusion case, which provides an 
explanation for Riggs' experimental results.  

There are to date very few direct dynamical numerical simulations of facilitated diffusion (for an up-to-date theoretical review see~\cite{Voituriez}). Ref.~\cite{Furini} simulated detailed protein-DNA interactions with molecular dynamics, focussing on a small, unconfined, DNA loop, while Refs.~\cite{Lang_I,Lang_II,Lang_III,Kafri1,Florescu} focused on few selected isotropic geometries and on a single protein diffusing inside the cell. Our aim here is to extend the study of facilitated diffusion in a number of ways, with a view to bridge the gap towards more realistic conditions. We start from the observation that DNA is strongly confined {\it in vivo}~\cite{jpcmreview}, and consider both isotropic and anisotropic containers, which are relevant for eukaryotic nuclei and bacterial cells respectively. Secondly, to assess the role of DNA conformations, we simulate different persistence lengths: this is important in eukaryotes as the persistence length of chromatin varies in the 40-200 nm range and is thought to be activity-dependent~\cite{jcb}. Finally, we extend our investigation to the case in which ${\cal O}(10)$ identical proteins simultaneously search for their target -- this is a more realistic scenario given e.g. the typical copy numbers of transcription factors inside a bacterium such as {\it E. coli}~\cite{Kafri2,Xie}. As we shall show, each of these realistic add-ons leads to important qualitative differences in the physics of facilitated diffusion.

Specifically, we find that the confining geometry and the DNA elasticity affect the mean search time in a major way: our simulations show that anisotropy slows down facilitated diffusion, while flexible conformations lead to a larger speed-up around the optimal sliding length with respect to semiflexible ones. These differences are much more pronounced for short chains.
Our main result is that considering a cell with multiple copies of a protein looking for their target, as is the case in nature, leads to a very different view on facilitated diffusion with respect to the standard one, based on single protein models. Instead of displaying a well-defined minimum as a function of the sliding length, the MST for multiple proteins (i.e. the time {\it one} protein first hits the target) is remarkably flat, so that facilitated diffusion offers a robust way to speed up the search, not relying on any optimisation linked to fine tuning of the parameters controlling the intracellular protein dynamics.
This trend holds for the chain lengths considered here, which are more than one order of magnitude larger than the sliding length: however, very long chains, which we have not simulated, may again display a different behaviour.

\section{Method}
Our approach is based on direct Monte-Carlo simulations of the whole facilitated diffusion process.
For the DNA, we use a coarse-grained bead-spring model where beads are joined by FENE springs and interact via a shifted and cut-off Lennard-Jones (LJ) potential. 
This provides a hard-core for the beads, hence capturing the polymer self-avoidance. The cut-off is at $2^{\frac{1}{6}}\sigma$, while the energy scale associated with the LJ potential is  $\epsilon=\frac{2}{5}k_B T$. 
Within this setting the natural length scale in our simulations is provided by the size of the 
chain beads, $\sigma$. We chose to simulate confining geometries with sizes of $\sim$ 10$\sigma$ - 30$\sigma$ as detailed below, and chains with contour length $L_c$ ranging from $60\sigma$ to $200\sigma$. 
Such relatively short polymers allow us to more stringently test the theories 
which typically rely on the existence of many unbinding and binding events and hence implicitly assume long polymers. 
Moreover conditions similar to our simulations may be designed experimentally by using 
naked DNA ($\sigma=2.5$ nm) of length 150-500 nm confined in vesicles of size ~50 nm -- a set-up which should be realisable with single molecule experiments {\it in vitro}. An intriguing alternative mapping is to yeast chromatin {\it in vivo}, where the appropriate coarse graining entails a value of $\sigma\sim 30$ nm (containing $\sim$ 3 kilo-base pairs of DNA), and typical chromosome length is $\sim 10^6$ base pairs (or $L_c \sim$ 300$\sigma$).

The elasticity of the DNA is modelled by a standard Kratky-Porod potential,
$U_b=-K_b\sum_i \vec{t}_i \cdot \vec{t}_{i+1}$, 
where $\vec{t}_i=\vec{r}_{i+1}-\vec{r}_i$, $\vec{r}_i$ being the position of the i-th monomer, and where the persistence length is determined in terms of $K_b$, the bending rigidity, via $l_p=K_b \sigma/k_B T$. We have considered the case of semiflexible ($l_p=20 \sigma$) and flexible ($K_b=0$) polymers, relevant to DNA and yeast active chromatin respectively~\cite{jcb}. 
Equilibrium DNA configurations are grown and relaxed inside cylindrical or spherical containers. For the confining geometry, we used cylinders having a total length of 30$\sigma$ and radii $R_{cyl}$ going from 4$\sigma$ to 10$\sigma$ to provide different aspect ratios for the container. 
Sphere radii $R_s$ range from 7$\sigma$ to 13$\sigma$ giving isotropic confining regions 
with the same volumes of the cylindrical counterparts. 
Rather than explicitly modelling polymer dynamics, the equilibrium DNA conformations 
are kept frozen during single 
diffusion processes. This quenched approximation for the DNA is in line with 
most previous theoretical and numerical work~\cite{Marko},~\cite{Lang_I} and 
can be justified by noting that the typical time needed by the polymer to change from one 
conformation to a completely uncorrelated one (Rouse time $\tau_R$~\cite{Rouse}) is $\sim L_c^2$, and for the cases we consider is larger than the timescale associated with protein diffusion, $\sim R^2/D_3$.

\begin{figure}
\label{fig:1}
\includegraphics[width=7.2cm]{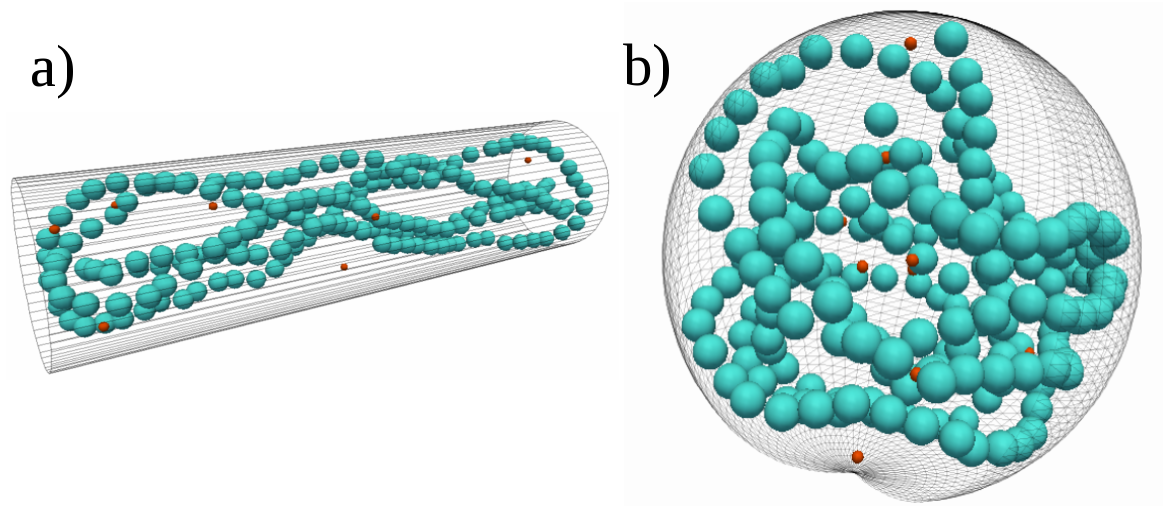}
\caption{(Color online) Some simulation snapshots. A semiflexible polymer is confined inside a cylinder (a) or a sphere (b), while some proteins (red dots) diffuse in 3D or slide along it.}
\end{figure}

Proteins are modelled as pointlike and non-mutually-interacting particles, whose initial 
positions are chosen to be uniformly distributed at the walls of the confining geometry. 
In the bulk, proteins diffuse with a coefficient $D_3=0.021 \sigma^2/$~ns. We use reflecting boundary conditions at the surface of the sphere or cylinder.
Interaction with DNA is modelled as follows: whenever the new proposed position for the protein falls inside one of the DNA beads, this starts sliding with probability $p$.
During each time step spent sliding, the protein may hop either to the right or to the left DNA bead, or detach from the genome with a probability $1-p$.
Different choices of $p$ correspond to different values of the sliding length $l_s$, the key optimisation parameter in the theory of facilitated diffusion. 
The duration of a sliding step, $\Delta t_s=48$~ns, was chosen in order to give a one-dimensional diffusion coefficient $D_1$ equal to $D_3$ in agreement with previous theoretical estimates~\cite{Marko} 
and simulations~\cite{Lang_I} (even though most experimental results suggest $D_1 < D_3$~\cite{Xie,D_3}). 
When the particle is at one of the two terminal beads of the chain it is 
always reflected, as in~\cite{Lang_I}. Some snapshots of our simulations are shown in Fig.~1a and 1b respectively for cylindrical and spherical confinement for semiflexible chains. 
The diffusion process for one protein stops as soon as it 
reaches the target, here modelled as an internal DNA bead. The interaction radius between the protein and the target, $r_b$ was equal to $r_b=0.35\sigma$ (which corresponds to about 1 nm for facilitated diffusion on naked DNA). 
For a fixed DNA equilibrium configuration the MST is estimated by averaging the 
time needed by one particle to reach the target over 100 initial conditions. A second 
average is then performed over $\sim 50$ different DNA configurations equilibrated in the 
confined region. 

\section{Results}

\subsection{Comparison with theoretical results}

\begin{figure}
\label{fig:2}
\includegraphics[width=6.7cm]{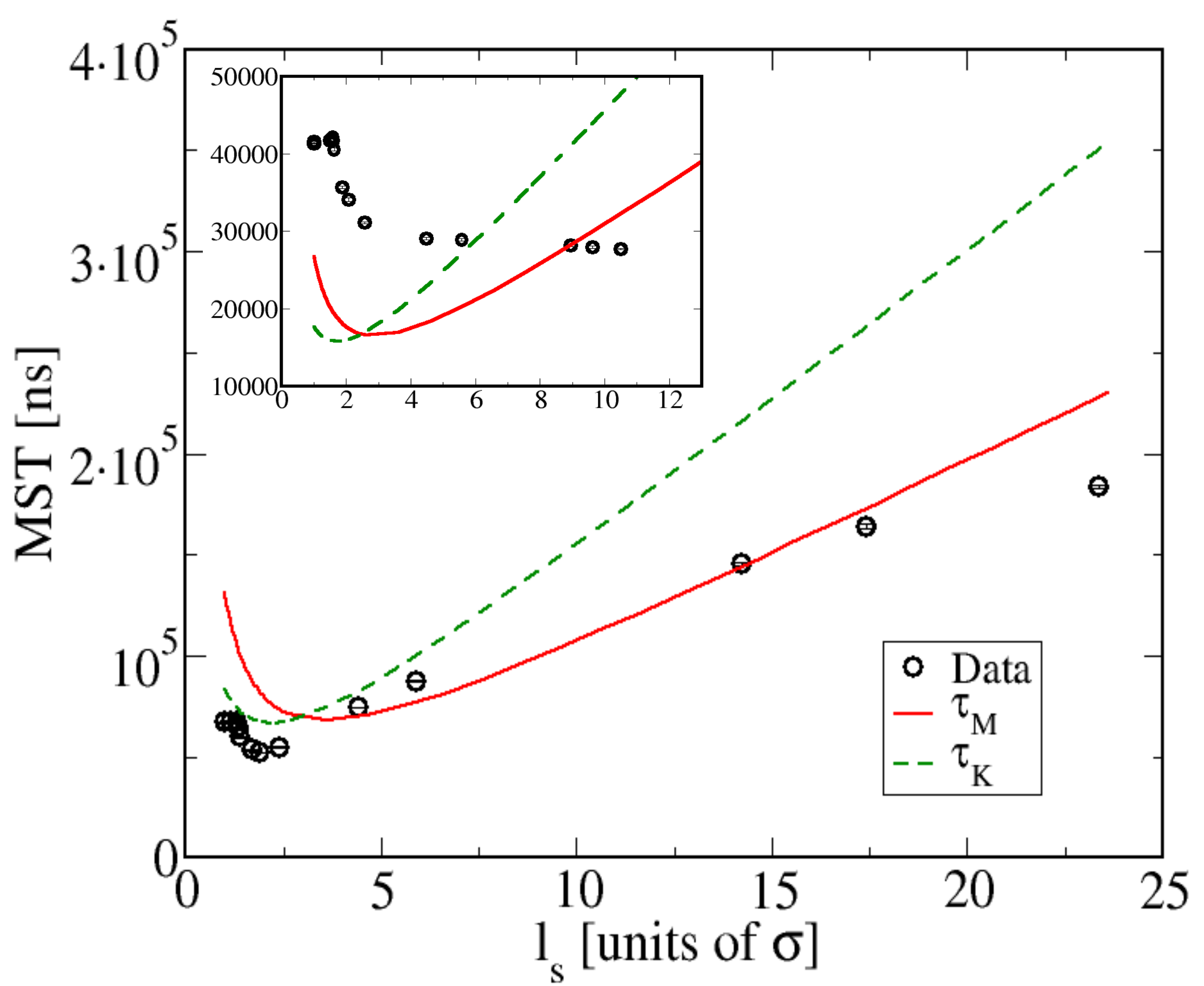}
\caption{(Color online) Dependence of the MST on $l_s$ is the case of a cylinder having radius $R_{cyl}=9\sigma$ and $L_c=200\sigma$ according to our data and to the theoretical predictions made in \cite{Marko} ($\tau_M$) and in \cite{Lang_I} ($\tau_K$). The case shown in the inset refers to short chains ($L_c=60\sigma$) and to very anisotropic confinement (cylinder with radius $R_{cyl}=4\sigma$ and side length
$L=30\sigma$).}
\end{figure}

We first compare our results to existing theoretical predictions by focusing on the 
cylindrical confining geometry and on the facilitated diffusion of a single protein.
Empty circles in Fig.~2 correspond to the simulated dependence of the MST on $l_s$ 
in the case of $l_p=20\sigma$, and $L_c=200\sigma$ (error bars are smaller than symbols). These points are compared with theoretical curves (solid lines) predicted in~\cite{Marko} ($\tau_M$) and in~\cite{Lang_I} ($\tau_K$). 
The latter gives a better estimate of the MST for small $l_s$, as well as of 
the optimal sliding length. For large $l_s$, according to both theoretical 
models, the MST should increase much faster than in our data, which instead tend to quickly saturate around a constant value. 
This discrepancy is more dramatic for shorter DNAs (see inset of Fig.~2, corresponding to a volume
fraction of $\sim 2\%$ -- a similar albeit smaller effect is observed for spherical confinement, discussed below)
where it is more likely for the diffusing protein to reach the target in just a few slides.
Such extreme events violate the hypothesis, used to estimate the MST theoretically, that the 
length scale over which the sliding mechanism occurs is much smaller than $L_c$~\cite{note}. 

\subsection{Search time for ten proteins}

\begin{figure}
\label{fig:3}
\includegraphics[width=6.7cm]{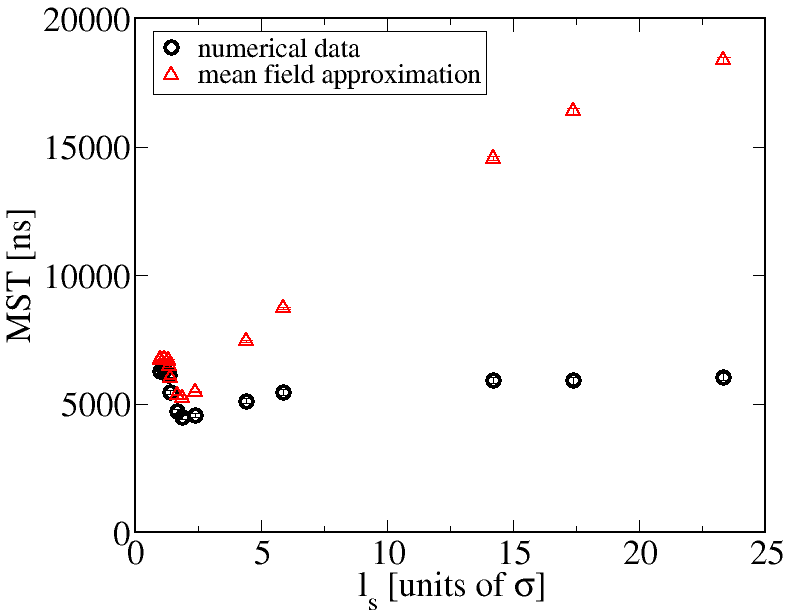}
\caption{(Color online) Comparison between the dependence on $l_s$ of the MST for one protein, scaled by 10 (mean field approximation) and for the first of ten particles (numerical data). Data refer to semiflexible DNA in a cylinder ($R_{cyl}=4\sigma$ and $L_c=200\sigma$).}
\end{figure}

So far we have considered the facilitated diffusion of a single protein. However in a cell there are typically more copies of the same protein which simultaneously search for the same target -- and a reaction is initiated once the first reaches it.

It is then natural to look at the time at which the first of several proteins finds the target. 
In Fig.~3 we plot the MST of the first of ten non-interacting and simultaneously diffusing particles -- for comparison we also show the corresponding MST curve for a single protein, rescaled by a factor of 10 as expected from a naive mean-field argument (critically analysed in Ref.~\cite{Kafri2}, see also below). The simulated data largely deviate from this simple estimate. 

\begin{figure}\label{fig:4}
\includegraphics[width=0.9\columnwidth]{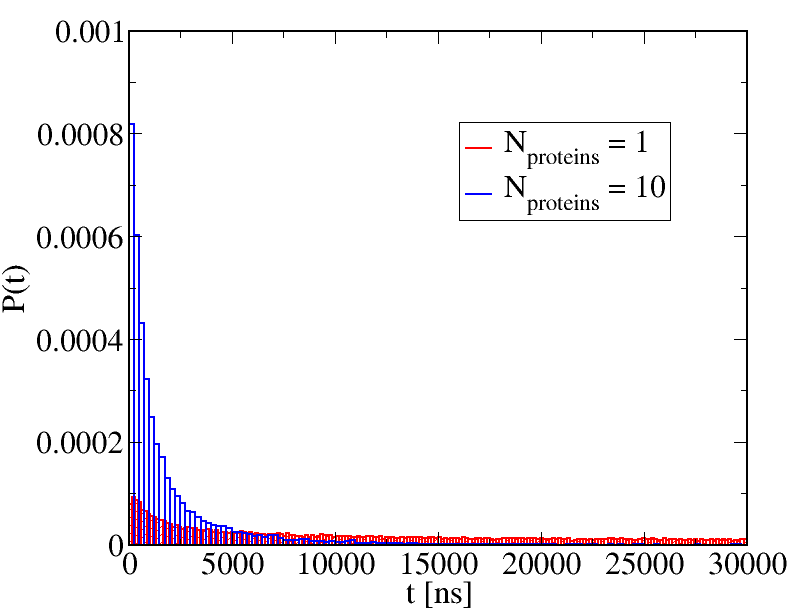}
\caption{(Color online) The normalised histogram of the number of particles 
which have reached the target at a given time $t$, 
in the case of $N_p=10$ (blue, top curve) and $N_p=1$ (red, bottom curve). 
Both graphs refer to a semiflexible polymer with $L_c=60\sigma$ and to a cylindrical confinement with  $R_{cyl}=4\sigma$. }
\end{figure}

The ultimate reason for this deviation lies in the markedly non-exponential nature of the search time distributions, which suggests the existence of multiple timescales in the dynamics.
This can be appreciated by plotting the distribution of arrival times, (i.e. the time to first reach the target), $P(t)$, shown in Fig.~4 for both the case of one single particle and for that of ten particles. 
From the linear-log plot in Fig.~5 it is apparent that neither distribution
is exponential -- a better fit is provided by a streched exponential
distribution (data not shown).

We note that, as our ten proteins are non-interacting, 
the probability that one of the
$N_p$ diffusing particles  reaches the target at time $t$, 
$R_{N_p}(t)=\int_0^t P(t^\prime)dt^\prime$, is given by
\begin{equation}
R_{N_p}(t)=1-(1-R(t))^{N_p},
\end{equation}
where $R(t)$ is the corresponding single protein probability 
(see e.g. \cite{Kafri1}). 
While directly considering the trajectories of 10 proteins in our
simulations, we have verified that this formula holds for our data.
If $P(t)$ were a simple exponential, this formula shows that there
would simply be a factor of 
$N_p$ between the MST of a single diffusing particle and the one where $N_p$ are
simultaneously diffusing in the system. 

\begin{figure}\label{fig:5}
\includegraphics[width=0.49\columnwidth]{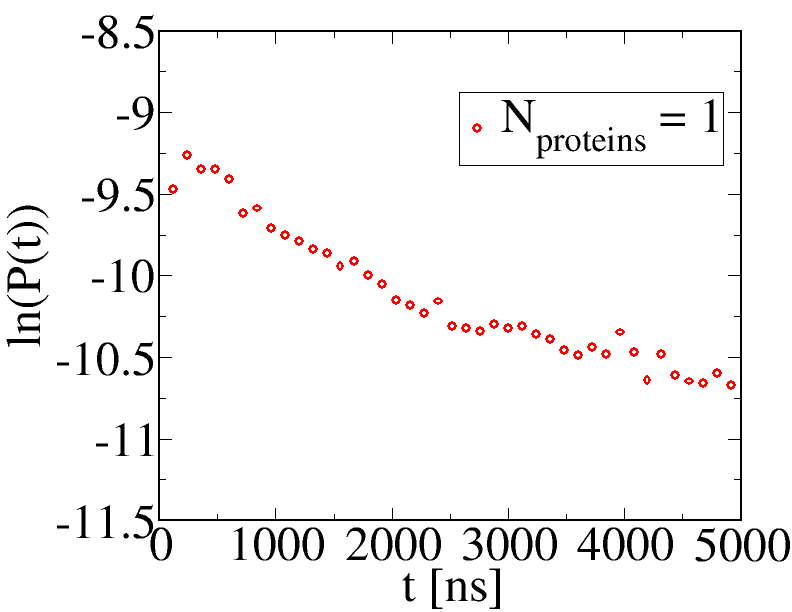}
\includegraphics[width=0.49\columnwidth]{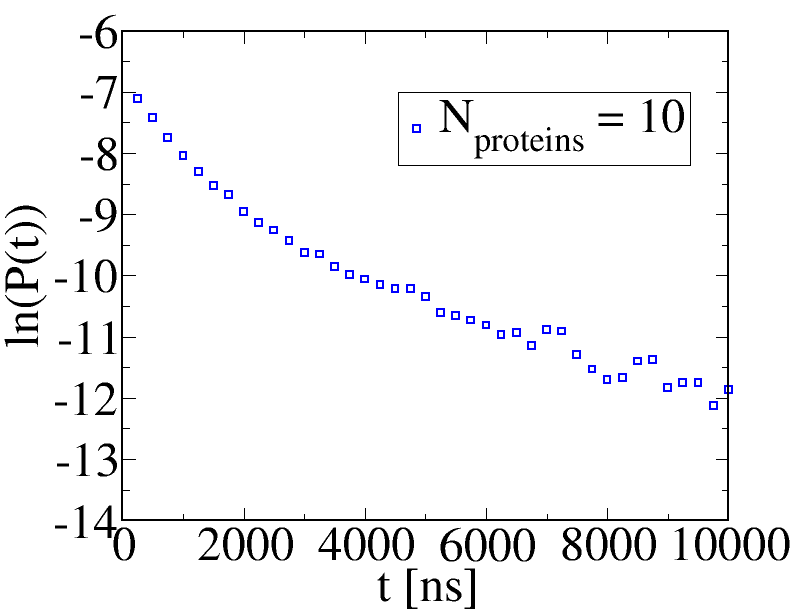}
\caption{(Color online) These graphs show the same data as Fig.~4 ($L_c=60\sigma$, semiflexible configurations confined in a cylinder with  $R_{cyl}=4\sigma$), but in a log-linear plot to stress the non exponential 
behavior of the distributions and they refer to the case of one particle (left) and of ten independent particles diffusing simultaneously (right).
}
\end{figure}

Importantly, we also find that the $l_s$ dependence of the MST of the first of ten proteins does {\em not} display a clear minimum anymore. This  suggests that, when more than one protein are searching for the target,  there is effectively no optimal sliding length! 
In other words, while facilitated diffusion is still faster than bulk diffusion, there is no need for the protein to fine tune the sliding length on the DNA to optimize the search process; rather, the advantages due to facilitated diffusion are robust and largely independent of the protein-DNA affinity. 
While this holds up to the longer chain length simulated here, we cannot rule out that very long chains may display a different behaviour. 

\subsection{Effect of confining geometry and DNA elasticity}

To further bridge the gap between simulations and biologically relevant conditions, it is important to address the effect of the aspect ratio of the confining geometry and of DNA elasticity. As mentioned in the introduction, there is a range of naturally occurring values for these parameters, and it is interesting to explore what impact tuning these has on the dynamics of facilitated diffusion. For instance, is there a difference in the efficiency of facilitated diffusion within eukaryotic nuclei, which are roughly spherical, and inside elongated bacterial cells? Do proteins search faster for a target inside euchromatin regions of the chromosomes, which are more flexible than inactive, heterochromatic ones~\cite{jcb}? To address these questions, we systematically analysed different confining conditions and polymer elasticities, and report below some typical results for the MST as a function of $l_s$.

\begin{figure}
\includegraphics[width=0.98\columnwidth]{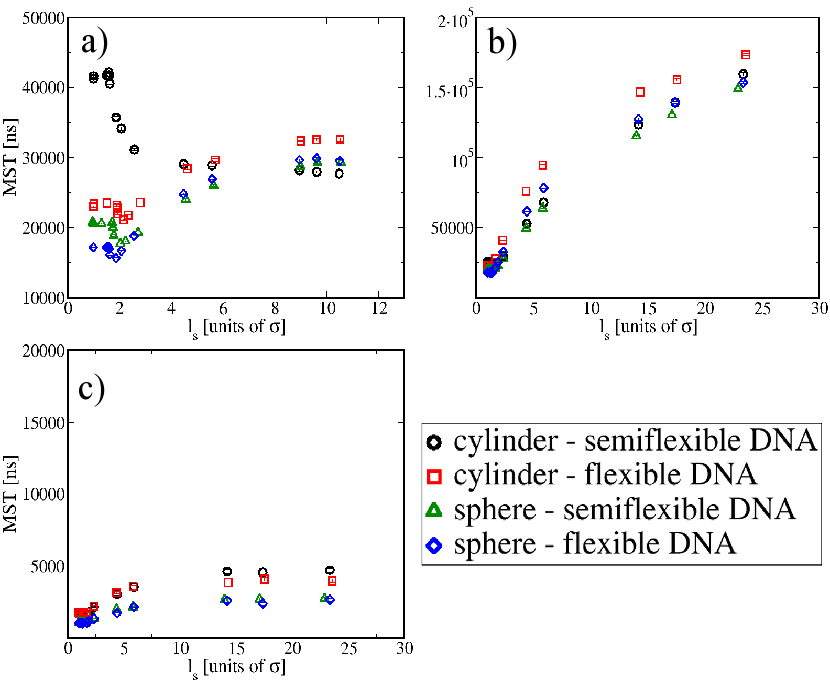}
\caption{(Color online) Dependence of the MST on $l_s$ for different geometries and DNA elasticity. Data refer to a single protein, $R_{cyl}=4\sigma$ and correspondingly $R_s=7\sigma$, when $L_c=60\sigma$ (a) and $L_c=200\sigma$ (b). (c) shows the MST for the first of ten particles for the same values of $R_s$ and $R_{cyl}$, when $L_c=200\sigma$.}
\label{fig:6}
\end{figure}

We have considered ten different cylindrical geometries, of aspect ratio $L/(2R_{\rm cyl})$ ($L$ and $R_{\rm cyl}$ are cylinder height and radius respectively, $L$ is kept fixed) variable between 3/2 and 15/4 -- we only report results on the MST for the most anisotropic case, as the difference with respect to the case of spherical confinement is more readily appreciable there. As anticipated, we also compare the cases of flexible ($l_p=0$) and semiflexible ($l_p=20$ $\sigma$) polymers.
In the Appendix we provide more details on the conformations of the polymer on which proteins perform their search for the target, and in particular discuss how they change with elasticity and confining geometry. Here we focus on the results for the MST.
Each panel of Fig.~\ref{fig:6} (from a to c) shows the $l_s$ dependence of the MST for four different cases: 
semiflexible and flexible DNA inside a cylinder, and semiflexible and flexible DNA inside a sphere  with the same volume. Note that to better highlight the effect of anisotropy on the MST, we focused on the most anisotropic cylindrical geometry, namely the one with aspect ratio $L/2R_{cyl}= 15/4$.

For the smallest chains considered ($L_c=60\sigma$, Fig.~6a) we observe major effects both of geometry and polymer elasticity, especially for small $l_s \stackrel{<}{\sim} 2\sigma$. 
In particular, in constrast to what happens when $l_p=20\sigma$ in a cylinder (Fig.~2 inset), in all other cases a well-defined optimal sliding length exists, and, more importantly, the MST is much smaller. This points out how both isotropy and flexibility can play an important role in enhancing the facilitated diffusion. Our results suggest that, at least for short chains, the search is faster within a sphere and on a flexible polymer.


Fig.~\ref{fig:6}b shows the same plots as in Fig.~\ref{fig:6}a, but relative to the case of $L_c=200\sigma$. 
Here the effects of the confining geometry and DNA conformations are less significant. 
This is due to the fact that the volume fraction occupied by the DNA now increases, so that even the stiffest conformations develop loops and local contacts, as they need to bend several times to fit in.  Moreover, long polymers are more homogeneously distributed in the cell than short ones, which renders the local environment of a DNA segment
less sensitive to the overall confining geometry, or polymer elasticity.
Finally, Fig.~\ref{fig:6}c shows the MST for the first of ten proteins to reach the target, for $L_c=200\sigma$. The data confirm the lack of an optimal sliding length. 
Interestingly, we also find that, as for the case of a single diffusing protein, the search is quicker within a sphere. The effect of flexibility is still present but milder than for a single protein.

An analysis of the statistics of 3D excursions in between slides, given below, sheds more light on these results. 

\subsection{Statistics of jumps and hops}

Here we provide an analysis of how the occurence probability of different 3D-diffusion mechanisms such as ``jumps'' and ``hops'' is affected by DNA elasticity and confinement geometry, with a view of explaining our results in Fig.~\ref{fig:6} on the influence of these factors on the MST.

``Hops'' and ``jumps''  are 3D-diffusion events that take place between two slides. In a hop the distance along the DNA between the take-off (detachment) and landing (re-attachment) point is ``small'', while in a jump it is ``large''. In our treatment we consider the distance ``small'' if it is less than 10 beads, and ``large'' otherwise.

\begin{figure}
\includegraphics[width=0.98\columnwidth]{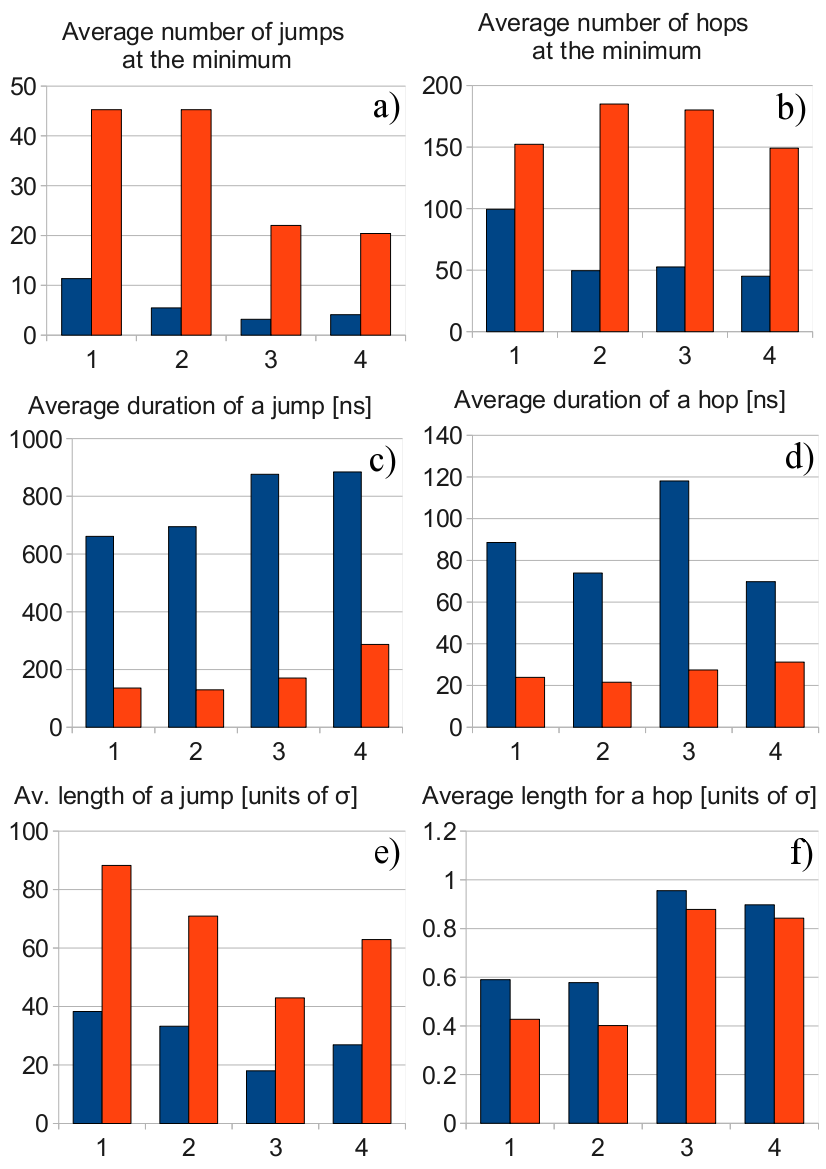}
\caption{(Color online) Histograms describing the average number (a, b) duration (c, d) and the average distance between the takeoff and landing points ('length', e, f) for jumps and hops. Blue (left) columns refer to $L_c=60\sigma$, while orange (right) columns refer to $L_c=200\sigma$. Numbers on the horizontal axis refer to different geometries and elasticities: (1) and (2) are for semiflexible DNA confined in a cylinder and in a sphere respectively, (3) and (4) refer to flexible DNA in a cylinder and in a sphere respectively. All data refer to the case of $R_{cyl}=4\sigma$ ($L/(2R_{cyl})=15/4$) -- and correspondently $R_s=7\sigma$.}
\label{fig:7:new}
\end{figure}

The key features of the statistics of jumps and hops are shown in Fig.~\ref{fig:7:new}.
The absolute numbers of jumps and hops give the clearest signal. As these quantities depend on the sliding length, we plot their value at the minimum MST (as a function of $l_s$) in Fig.~\ref{fig:7:new}a,b. For $L_c=60\sigma$, both quantities are largest in the case of a semiflexible DNA in cylindrical confinement, which correlates well to a larger MST in this situation. Furthermore, still for $L_c=60\sigma$, in the case of spherical confinement a smaller number of jumps is required on a flexible DNA (case 4 in Fig.~\ref{fig:7:new}) than on a semiflexible one (case 2 in Fig.~\ref{fig:7:new}), to reach the target, again in agreement with the MST trends in Fig.~\ref{fig:6}. Facilitated diffusion on a flexible DNA inside a cylindrical container would require the smallest number of jumps, which does not match the fact that the MST in this case is not optimal. This is because in this situation the number and duration of hops also plays a role: these are relatively numerous, and, more importantly, each of these lasts about $50\%$ longer, with respect to the average of the other cases (see Fig.~\ref{fig:7:new}d). 

It is also interesting to monitor the ratio between the number of jumps and the number of hops, which instead does {\it not} depend on the sliding length. This ratio is once again affected by the polymer elasticity, more so in the case of a cylindrical confinement, where for a semiflexible DNA it is equal to 0.11, almost twice as large as for a flexible DNA (0.06). In the case of spherical confinement the ratio between hops and jumps is again equal to 0.11 for semiflexible polymers, and rises to 0.09 for flexible ones.
These differences are related to the average backbone distance between the takeoff and landing points, in both jumps and hops: as can be seen in Fig.~\ref{fig:7:new}e and f, particles perform longer jumps along a semiflexible than along a flexible polymer. In contrast, hops are shorter when performed along semiflexible configurations. This is evidence of the fact that stiffness favours either very short or very long range (in terms of chemical distance) three-dimensional displacements, while flexibility allows intermediate-range displacements, due to enhanced local bending. Short-range 3D displacements are advantageous for short DNA chains, as witnessed by our MST data in Fig.~\ref{fig:6}.

When longer polymers are considered ($L_c=200\sigma$), though, the trends are less clearcut. 
On the one hand, for very low values of $l_s$, the MST in a cylinder is again larger with respect to that in a sphere, and once more this is in line with the trends in the total number of jumps.
On the other hand, now the semiflexible DNA inside cylindrical confinement no longer gives the slowest search, especially for larger values of $l_s$, where the MST is largest for a flexible chain inside a cylinder. 
This is in part because the frequent intermediate-range displacements performed on flexible polymers confined in cylinders make a less efficient strategy in longer chains, as this wastes time in sampling nearby (in terms of chemical distance) DNA regions. 
At the same time, it is interesting to note that jumps always take (slightly) longer along flexible than along semiflexible chains. We ascribe this to the fact that in flexible polymers most of the nearby beads are also adjacent along the chain, and it is less likely for a protein to find a closeby bead which is very far along the DNA. This is a disadvantage especially in the case of long polymers, where jumps are expected to play an important role in favouring a better sampling of DNA.
As a result the MST inside a cylinder is slightly longer in the flexible case.

We finally note that in all cases isotropic confinement reduces the MST, for both small and long polymers. This can be explained by noting that in a sphere any two points are on average closer together than inside a cylinder. This is quantified via  Fig.~11 (see Appendix), showing the mutual distance distribution function, $\Pi(r)$~\cite{note2}.
This distribution is much more spread for cylinders than for spheres, which allows a particle in a sphere to reach its target after a considerably smaller number of 3D-diffusion events.

 

To summarise, in this section we have analysed the statistics of jumps and hops with a view of understanding more deeply the results in Fig.~\ref{fig:6} on the effect of geometry and elasticity on the MST trends. 
Firstly, we have seen that, especially for short polymers, the MST trends can be rationalised by looking at the total number of jumps and hops needed to get to the target. This may be viewed as a consequence of the fact that the mutual distance distribution function goes to  zero at a much shorter distance in an isotropic geometry.
Secondly, we have shown that in a cylinder, 3D-displacements along a flexible chain usually take longer than in the presence of semiflexible DNA. This increases the MST of long polymers, where the search requires several such jumps.

\section{Summary and Conclusions}

In summary, we have presented detailed Monte Carlo direct dynamical simulations of the facilitated 
diffusion of one or several proteins inside a cylindrical or spherical cell in presence of 
equilibrated DNA configurations. We have compared our simulation results with theoretical 
predictions for the mean search time, the time needed for one protein to first reach its target, for instance a promoter along the genome. 
For short DNA chains we find a large discrepancy, which we attribute to the fact that proteins can reach the target with no or few sliding steps 
along the genome. This discrepancy attenuates as the contour length of the DNA increases.
Our main result is that the average search time needed for the first of several copies of the same protein looking for their target simultaneously shows a qualitatively different dependence on physical parameters: most notably, there is no longer an optimal sliding length leading to the fastest search. 
This is important biologically, as in bacterial or eukaryotic cells the typical copy numbers of e.g. transcription factors is low but larger than unity (typically $\sim$10~\cite{Xie}). 
This fact may then be interpreted as a way to ensure that the search for the specific target of a protein is not crucially dependent on the precise value of its non-specific affinity with DNA. 
Interestingly, we have also found that polymer elasticity and the confining geometry may both affect the mean search time: facilitated diffusion is most effective inside spherical cells and in presence of flexible polymers. 
We find it intriguing to note that these advantages may be exploited for facilitated diffusion inside the spherical nuclei of eukaryotic cells such as yeast, where an enrichment in local gene activity is thought to correlate with an enhanced chromatin flexibility~\cite{jcb}. Finally, we hope that our results may stimulate {\it in vitro} single molecule experiments on facilitated diffusion of proteins on DNA molecules of different contour lengths and under different confining geometries.

\section{Acknowledgements} 
The work was carried out under the HPC-EUROPA2 project (project number: 228398) with the support of the European Commission - Capacities Area - Research Infrastructures.

\appendix

\section{Characterisation of DNA conformations}

In this Appendix we give more details on the DNA conformations, and on their
dependence on the polymer flexibility and on the confining geometry. 

\begin{figure}\label{fig:8}
\includegraphics[width=0.49\columnwidth]{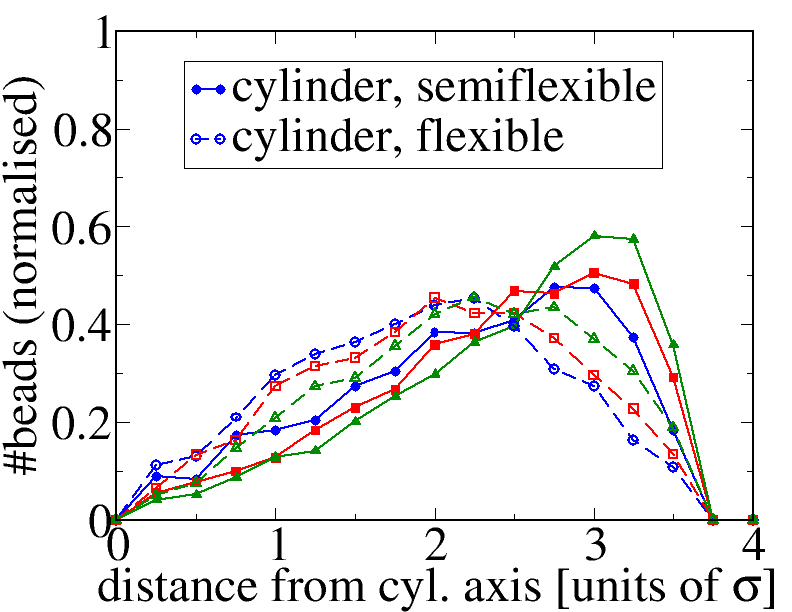}
\includegraphics[width=0.49\columnwidth]{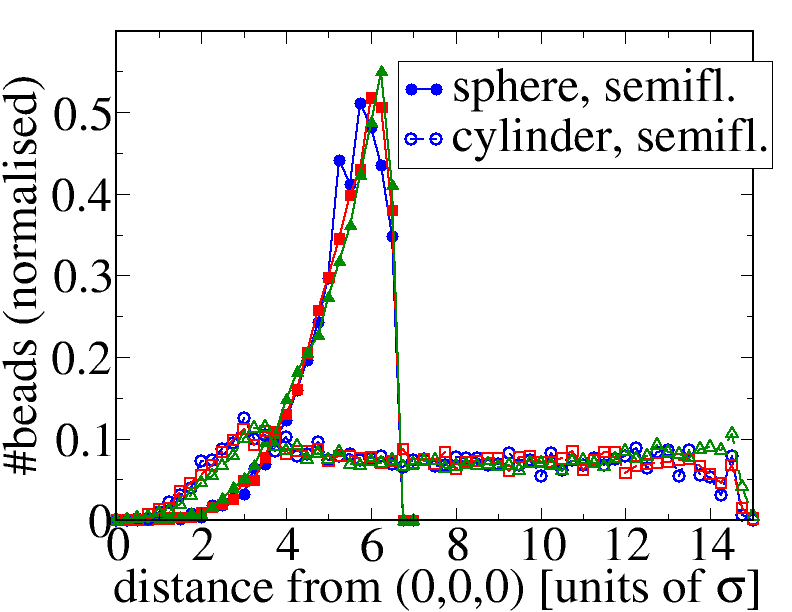}
\caption{(Color online) Distribution of the distance of beads from the cylinder axis for semiflexible and flexible DNA (left) and of the distance from the cylinder and sphere centre for semiflexible DNA (right). Different colours (symbols) refer to different contour lengths $L_c=60\sigma$ (blue, circles), $120\sigma$ (red, squares) and $200\sigma$ (green, triangles). Both graphs refer to $R_{cyl}=4\sigma$, and correspondently to $R_s=7\sigma$.}
\end{figure} 

Fig.~8 (left) shows the distribution of the distance of DNA beads from the cylinder axis. Semiflexible polymers tend to concentrate more near the walls, since this allows them to minimise the bending. Flexible polymers instead are more homogeneously distributed inside the confining geometry. 
Fig.~8 (right) shows the distribution of the distance of the beads of a semiflexible DNA from the centre of the confining geometry (either a cylinder or a sphere), in the case of the most anisotropic cylindrical geometry ($L/(2R_{cyl})=15/4$) and of a sphere having the same volume ($R_s=7\sigma$). While in the case of a cylinder all distances between $R_{cyl}=4\sigma$ and $L/2=15\sigma$ are almost equally likely, in the sphere isotropy leads to a peak at distances slightly smaller than $R_s$. [Similar results are obtained when comparing the distribution of beads for semiflexible and flexible DNA in a sphere or for flexible DNA in different geometries.]


\begin{figure}\label{fig:9}
\includegraphics[width=0.98\columnwidth]{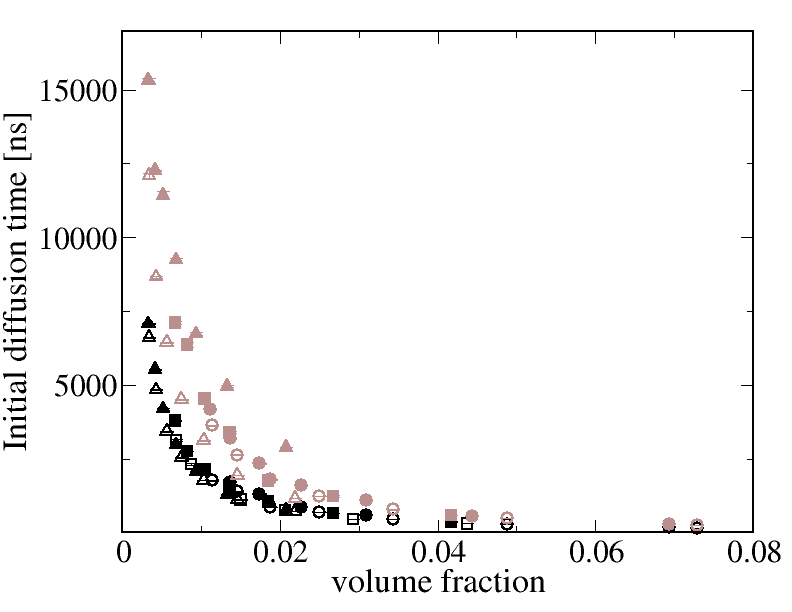}
\caption{(Color online) Dependence of the initial diffusion time, spent on average by one particle in performing 3D-diffusion in the bulk before starting interacting with DNA, as a function of the volume fraction. Different colours (shades of gray) refer to different confining geometries and elasticities: 
semiflexible (black) and flexible (light gray) polymers inside a cylinder (filled symbols) and a sphere (open symbols). Values for different DNA contour lengths are shown: $L_c=60\sigma$ (triangles), $L_c=120\sigma$ (squares) and $L_c=200\sigma$ (circles).}
\end{figure}

The spatial distribution of the DNA correlates well with the ``initial diffusion time'', i.e. the time that a protein, 
starting from the surface of the cell (whether spherical or cylindrical in shape) takes before ``touching'' the DNA for the first time, independently of whether or not a slide is performed then. This initial diffusion time, computed for the different DNA conformations considered, is shown as a function of the volume fraction in Fig.~9. 
This quantity is larger in the case of flexible polymers, and it is maximal for short flexible chains, in which case the initial diffusion time impacts considerably on the total MST. 
This trend, however interesting, is not impacting much on the MST, whose dependence on geometry and flexibility is more crucially determined by the average number of jumps and hops (see main text).

\begin{figure}\label{fig:10}
\includegraphics[width=0.98\columnwidth]{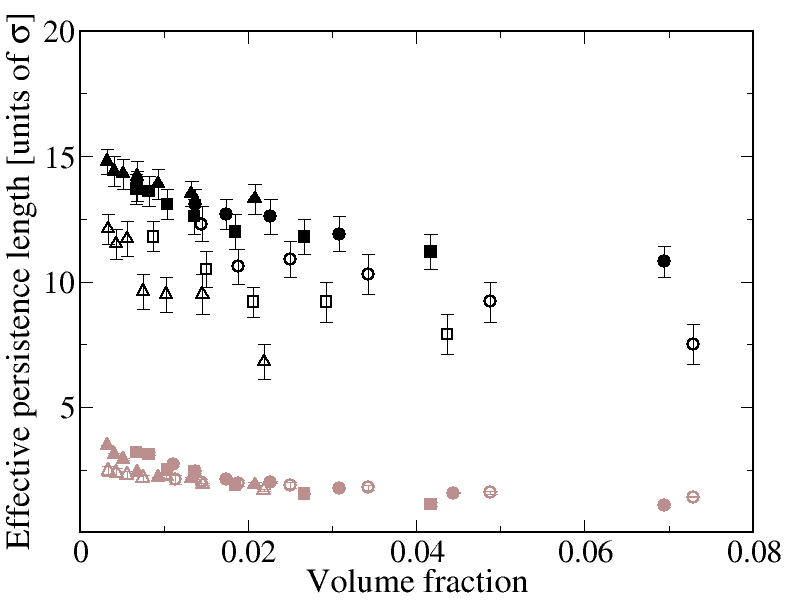}
\caption{(Color online) The graph shows the dependence of the persistence length on the volume fraction in the case of $L_c=200\sigma$ and $R_{cyl}=4\sigma$ for semiflexible DNA confined in a cylinder. Different colours (shades of gray) refer to diffent confining geometries and elasticities: 
semiflexible (black) and flexible (light gray) polymers inside a cylinder (filled symbols) and a sphere (open symbols).
Values for different DNA contour lengths are shown: $L_c=60\sigma$ (triangles), $L_c=120\sigma$ (squares) and $L_c=200\sigma$ (circles).}
\end{figure}

A more detailed characterisation of DNA elasticity is provided by the tangent-tangent (normalised) correlation function $C_t(m)=\langle \vec{t}_i \cdot \vec{t}_{i+m}\rangle/|\vec{t}_i||\vec{t}_{i+m}|$, where $\vec{t}_i=\vec{r}_{i+1}-\vec{r}_i$, and $\vec{r}_i$ gives the position of the $i$-th bead. The characteristic decay length of $C_t(m)$ (which is defined by assuming an exponential decay) defines the persistence length. In the unconfined case, for semiflexible polymers this is $l_p = 20\sigma$ as expected. When DNA is confined, though, it needs to bend several times to fit in the confining geometry, and $C_t$ consequently decreases more rapidly in space, so that the measured \emph{effective} persistence length, $l_{p,eff}$, defined according to $C_t(m) \sim e^{-\frac{m}{l_{p,eff}}}$ is always smaller than $l_p$.
Fig.~10 shows $l_{p,eff}$ as a function of the volume fraction in the four different cases of a semiflexible and of a flexible polymer in a cylinder and in a sphere. It is interesting to notice here that the isotropic confinement of the sphere determines a smaller $l_{p,eff}$ with respect to the one measured in a cylinder, for semiflexible configurations with the same contour length. This is due to the fact that the DNA, in order to fit into a  sphere,  needs to bend more than in a cylinder with the same volume.

Finally, an important DNA configurational property affecting the MST is given by  the ``local environment'' around a given bead of the chain that can be probed, for example, by the distribution probability of all mutual distances (between any two beads in the chain), which we call $\Pi(r)$. This is plotted in Fig.~11, only for beads whose distance along the chain is larger than $5\sigma$ (this threshold is chosen to filter out the information from consecutive beads  along the chain). Fig.~11 shows that for $L_c=60\sigma$ the typical distance between two beads is larger in a cylinder, for semiflexible chains. This is because flexible chains are more compact on average so that points are on average closer together. Furthermore, in a sphere $\Pi(r)$ goes to zero at a smaller value of the interparticle distance since, despite having the same volume, cylinders are more elongated and therefore allow any two beads to be farther apart with respect to what happens in spheres, where the maximum distance between two points is set by the diameter.

 \begin{figure}\label{fig:11}
\includegraphics[width=0.98\columnwidth]{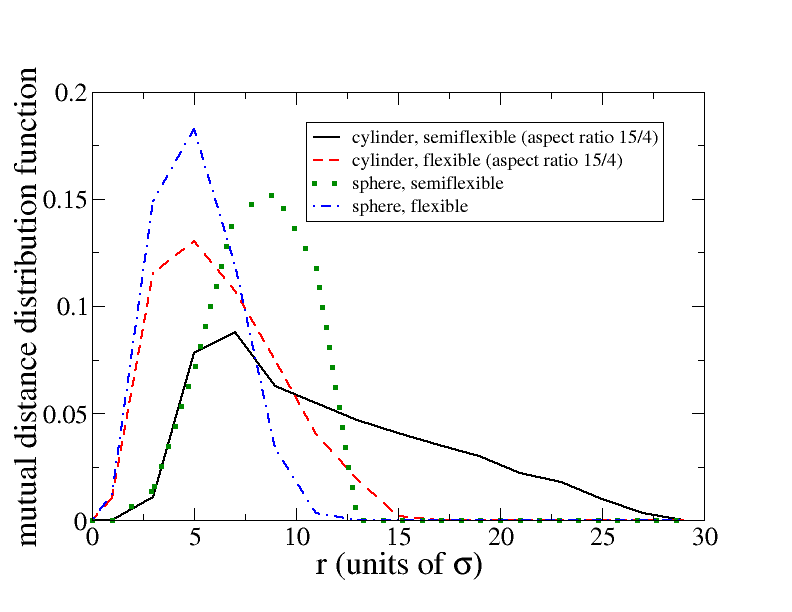}
\caption{(Color online) Dependence of the mutual distance distribution function, $\Pi(r)$ for different confinement geometry and polymer elasticity, see legend. Data correspond to $L_c=60\sigma$, to $R_{cyl}=4\sigma$ and to $R_s=7\sigma$. In the calculation only beads separated by $5\sigma$ or more along the backbone of the polymer are included.}
\end{figure}


\begin{thebibliography}{99}
\bibitem{Alberts} B.~Alberts {\it et al.}, {\it Molecular Biology of the Cell}, Garland Science, New York (2002).
\bibitem{Voituriez} M.~Sheinman, O. Benichou, Y. Kafri and R. Voituriez, 
 {\it Rep. Prog. Phys.} {\bf 75}, 026601 (2012).
\bibitem{Voituriez2} S. Condamin, 
O. Benichou, V. Tejedor, R. Voituriez and J. Klafter,
{\it Nature} {\bf 450}, 77 (2007); O. Benichou, C. Chevalier, B. Meyer and R. Voituriez,
{\it Phys. Rev. Lett.} {\bf 106}, 038102 (2011).
\bibitem{Riggs} A.~D.~Riggs, S.~Bourgeois, M.~Cohn, {\it J. Mol. Biol.} {\bf 53}, 401 (1970).
\bibitem{Berg_I} P.~H.~von Hippel, O.~G.~Berg, {\it J. Biol. Chem.} {\bf 264}, 675 (1989).
\bibitem{metzler} R. van den Broek, M.~A. Lomholt, S.~M.~J. Kalitz, R. ~Metzler, G.~J.~L. Wuite, 
{\it Proc. Natl. Acad. Sci. USA} {\bf 105}, 15738 (2008).
\bibitem{def_jump_hop} We discriminate between two kinds of 3D excursions in between two slides, ``hops'' or ``jumps'', according to the distance between the detachment and reattachment points along the chain. In hops both these should be below some threshold (see the Section IIID).
\bibitem{Marko} S.~E.~Halford, J.~F.~Marko, {\it Nucleic Acids Res.} {\bf 32}, 3040 (2004).
\bibitem{Furini} S.~Furini, C.~Domene, S.~Cavalcanti, {\it J. Chem. Phys. B} {\bf 114}, 2238 (2010).
\bibitem{Lang_I} K.~V.~Klenin, H.~Merlitz J.~Langowski, {\it Phys. Rev. Lett.} {\bf 96}, 018104 (2006);
\bibitem{Lang_II} 
H.~Merlitz, K.~V.~Klenin J.~Langowski, {\it J. Chem. Phys.} {\bf 124}, 134908 (2006).
\bibitem{Lang_III} H.~Merlitz, K.~V.~Klenin and J.~Langowski, 
{\it J. Chem. Phys.} {\bf 125}, 041906 (2006).
\bibitem{Kafri1} M.~Sheinman, Y.~Kafri, {\it Phys. Biol.} {\bf 6}, 0160033 (2009).
\bibitem{Florescu} A.-M. Florescu, M. Joyeux, {\it J. Chem. Phys.} {\bf 130},
015103 (2009).
\bibitem{jpcmreview} D. Marenduzzo, C. Micheletti, E. Orlandini,
{\it J. Phys.: Condens. Matt.} {\bf 22}, 283102 (2010).
\bibitem{jcb} P.R.Cook, D.Marenduzzo, {\it J. Cell. Biol.} {\bf 186}, 825 (2009).
\bibitem{Kafri2}
O.~Benichou, Y.~Kafri, M.~Sheinman and R.~Voituriez, 
{\it Phys. Rev. Lett.} {\bf 103}, 138102 (2009).
\bibitem{Xie} X.~S.~Xie, P.~J.~Choi, G.~Li, N.~K.~Lie and G.~Lia, 
{\it Ann. Rev. Biophys.} {\bf 37},417 (2008).
\bibitem{Rouse} P.~E.~Rouse, {\it J. Chem. Phys.} {\bf 21},(1953).
\bibitem{D_3} M.~B.~Elowitz, M.~G.~Surette, P.~Wolf, J.~B.~Stock, and S.~Leiber, 
{\it J. Bacteriol.} {\bf 181}, 197 (1999).
\bibitem{note}No deviation was seen in~\cite{Lang_I,Kafri2}, 
where typically sliding lengths were $\ll$ the DNA contour length. 
\bibitem{note2} $\Pi(r)$ is found by constructing the normalised histograms of the number of particles which are at a mutual distance between $r$ and $r+dr$.

\end{thebibliography}
\end{document}